# Structure/property relationship of semi-crystalline polymer during tensile deformation: A molecular dynamics approach

Cheng Li, Yingrui Shang*, Huan Xu, Jinqing Li, Shichun Jiang**
College of Material Science and Engineering, Tianjin University, Tianjin, 300072
*yrshang@tju.edu.cn **scjiang@tju.edu.cn

## Abstract

A coarse-grained molecular dynamics model of linear polyethylene-like polymer chain system was built to investigate the responds of structure and mechanical properties during uniaxial deformation. The influence of chain length, temperature, and strain rate were studied. The molecular dynamic tests showed that yielding may governed by different mechanisms at temperatures above and below $T_g$. Melt-recrystallization was observed at higher temperature, and destruction of crystal structures was observed at lower temperatures beyond yield point. While the higher temperature and lower strain rate have similar effects on mechanical properties. The correlated influences of time and temperature in the microscopic structures are more complicated. The evolution of microscopic characteristics such as the orientation parameter, the bond length, and the content of trans-trans conformation were calculated from the simulation. The results showed that the temperature have double effects on polymer chains. Higher temperature on one hand makes the chains more flexible, while on the other hand shortens the relaxation time of polymers. It is the interaction of these two aspects that determine the orientation parameter. During deformation, the trans conformation has experienced a rising process after the first drop process. And these microscopic structure parameters exhibit critical transaction, which are closely related to the yield point. A hypothetical model was thus proposed to describe the micro-structure and property relations based the investigations of this study.

# Introduction

Semi-crystalline polymer materials pose long standing puzzles in its structure/property relations, mainly due to the hierarchical structures of polymer crystalline and the coexistence of amorphous and crystalline domains. Moreover, various thermal processes[1-4]through practices such as extrusion, injection, compression, or annealing, may introduce significant differences in morphology evolution of this amorphous-crystalline binary system. The structure and property relations of semi-crystalline polymers are heavily affected by the characteristic of polymer chains. The high molecular weight and long relaxation time of macromolecules bring up the complexity into the structure/property relations.

The interests on structure/property relation of semi-crystalline polymers have barely fade since its discovery in 1960s, mostly because of its wide usage in industry and excellent cost performance ratio.[5],[6]The experiment methods such as X-ray synchrotron, infrared spectroscopy, differential scanning calorimetry, and scanning electron microscopy are widely used to investigate the structural evolution of polymer materials. The classical Peterlin's model[7] proposed the orientation and fracture of spherulites and lamellar, and formation of microfibril structures during deformation of semi-crystalline polymer. Juska and Harrison hypothesized a melt-recrystallization procedural.[8] While more and more recent studies suggested that, besides the structural evolution of the crystalline domain, the amorphous part plays an important role during deformation process.[9-12]

Experimental techniques such as infrared spectroscopy(IR),small-angle X-ray scattering (SAXS), wide-angle X-ray scattering (WAXS), atomic force microscopy (AFM), and diffraction scanning calorimetry have been employed and new discoveries and theories have been put forward for recent decades. Feng Zuo, Benjamin S. Hsiao et al.[13] have investigated isotactic polypropylene (iPP) deformation with *in situ* SAXS, and WAXD. It was observed that at room temperature the distraction of lamellar crystals is dominant while at higher temperature (>60$^{o}$C) the formation of oriented folded chain crystal lamellar is dominating. And this phenomenon is attributed to the chain entanglement and tie chains between crystal lamellar and the relative strength of amorphous part to the crystalline domain at different temperatures. Yongfeng Men, Gert Strobl et al[14] investigated the interplay of the amorphous and crystal blocks in semi-crystalline polymer during deformation, and found that the state of the amorphous part and the stability of the crystal block act together to determine the critical strain (yield strain). While accordingly, the tie chains are of lesser importance compared to the state of amorphous domain as a whole. The experiment methods have played an irreplaceable role in scientific research on these issues, these measurements can provide partial or statistical structural information of the material, however, this is insufficient for revealing the complex micro-/meso-structure property relations, some important information is still missing.

Nevertheless, it is difficult to investigate the micro-/meso-structures closely with conventional experimental measurements, and studying their influences on macroscopic properties in real time has been always a challenge. The molecular dynamics (MD) simulation provides a new route to reveal the details in structure/property relations in small scales. With MD simulation, *in situ* study can be readily conducted in chain configuration/conformation as well as mesoscale structures. The results of simulation and experiment investigations can be compared to help understand the essence of how synthesis and processing of polymer materials may determine the mechanical properties. Recent development in hardware and algorithms makes it possible to simulate large scale polymer system within reasonable CPU time. Significant progress has been made in exploring the microstructure/property relationships of polymer materials through MD simulation[15-17]. Takashi Yamamoto[18]studied polyethylene with a united atom model in fiber formation and large deformation by MD simulations. The study compared the structure transformation in fiber axis with transverse direction. The deformation along the fiber axis was almost linear and elastic before yielding, and caused large reorientation of the tilted chains in the crystals. After yielding, cavitation was occurred in amorphous regions. While along the transverse direction, the molecular chains give rise to the 90 ° reorientation toward the uniaxial deformation direction, breaking and reformation of the crystalline texture was also emerged, simultaneously. Recently, In-Chul Yeh, Gregory C. Rutledge, et al.[19] through MD model to investigate deformation of the semi-crystalline polyethylene at different strain rates and temperatures. It was learned that cavitations emerged at low temperature or high strain rate. While at higher temperature or lower strain rate the melt/recrystallization phenomenon was be observed. The results exhibit that the interaction of crystalline and noncrystalline domains is a crucial factor in determining the mechanical properties during tensile deformation. The interface Monte Carlo (IMC) method was proposed and employed to prepare PE model with coexisting amorphous and crystalline domain.

The purpose of this paper is to reveal the structure/property relations of semi-crystalline polymers, with MD simulation on the uniaxial tensile deformation process under various strain rates and temperatures. The model with coexisting amorphous and crystalline domains was established though isothermal crystallization process. This model is thermally stable and more realistic comparing to other related works such as the IMC method. The micro-/meso-structure evolution during deformation was investigated in details. And hypothetical mechanism was proposed to describe the influence of determinate structural characteristics on mechanical properties.

## Numerical Model

### Coarse graining

The simulation tests run on an in house workstation SP2EHIEQ in parallel mode, with 14 CPUs and 28 threads. In this paper a linear polyethylene-like molecular was chose as the study object. Through coarse-graining, the carbon-hydrogen bonds are ignored. A coarse-grained bead represents one monomeric unit which is connected to neighbor beads by harmonic springs. With this simplification, the angle in this system actually represents the torsional angle in the atomic backbones. This coarse-graining method was proposed by Meyer and Müller,[20],[21] and has been applied widely.[22],[23] The mass of one bead equals to the total mass of the monomeric unit.

**Force field and related parameters**

The force field is composed of bond stretching potential, angular bending potential, and nonbond interaction potential. No charges and torsional potential was considered in this model. Thus, the total potential of the system can be described as follows.

$$E_{total} = E_{bond} + E_{angle} + E_{pair} \tag{1}$$

A harmonic form of the bond potential is adopted.

$$E_{bond} = \frac{K_{bond}}{2}\left(r - b_0\right)^2 \tag{2}$$

where $b_0$ is the equilibrium bond length, $K_{bond}$ is the force constant, which determined the bond stiffness under extension.

The angle bending potential which contains information on the torsional states of the atomistic backbone was derived directly from the Boltzmann-inverted angle distribution of the atomistic trajectories. The angle potential is exhibited in Fig 1. Three minima at 95°, 126°, 180° are displayed, corresponding to gauche-gauche, trans-gauche, trans-trans conformations of the atomic backbone chain, respectively.

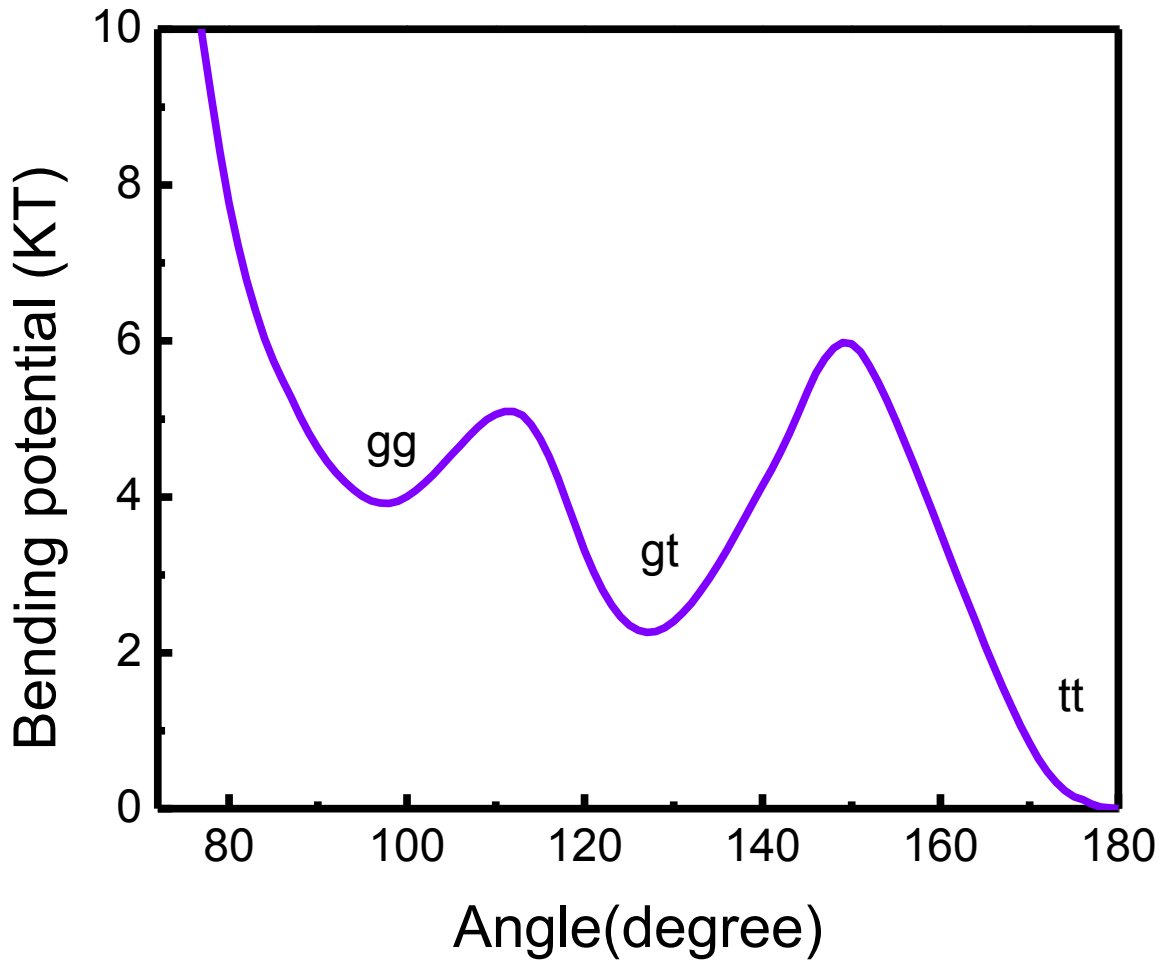

Fig 1. Bending potential of CG-PE model.

Pair potential is defined as the potential that between pairs of atoms within a cutoff distance and the set of active interactions typically changes over time. In this simulation the non-bonded interaction potential was adopted a Lennared-Jones 9-6 potential.

$$E_{pair} = 4\varepsilon_0\left(\left(\frac{\sigma_0}{r}\right)^9 - \left(\frac{\sigma_0}{r}\right)^6\right), r < r_{cut} \tag{3}$$

where $\varepsilon_0$ is a parameter that determines the depth of the potential well on the equilibrium position. $\sigma_0$ represents the equilibrium distance of a pair of beads. The cutoff distance $r_{cut}$ is estimated from the equilibrium distance of the potential as $r_e = \left(\frac{3}{2}\right)^{1/3}\sigma_0$, beyond which the van der Waals interaction between the coarse-grained particles is omitted.

The values of miscellaneous modeling coefficients and deformation conditions are listed in Table 1.

Table 1. Values of coefficients.

| Description | Implemented values |
|---|---|
| Temperature | 110-550K |
| Strain rate† | $1.6149\times10^{-6}$-$3.2298\times10^{-5}$nm/ps |
| $K_{bond}$ | $3.7968\times10^{-17}$J/nm$^2$ |
| $b_0$ | 0.26nm |
| $\varepsilon_0$ | $2.8666\times10^{-21}$J |
| $\sigma_0$ | 0.4638nm |
| $r_{cut}$ | 0.5304nm |
| $\tau$ | 1.61ps |
| $m$ | 28.04g/mol |

† Engineering strain is used in this paper.

In this simulation a nondimentionalized unit system is used. The nondimension factors are listed in Table 2.

Table 2 Nondimentionalization

| ND factors | Implemented values |
|---|---|
| $\varepsilon_0^*$ | $7.5936\times10^{-21}$J |
| $\sigma_0^*$ | 0.5200nm |
| $m^*$ | 28.04g/mol |
| $k_B^*$ | 1.3806J/K |

Other unitless parameter values are thus derived accordingly. In this nondimentionalization system, T=1 corresponds to the temperature about 550K. The time step of 0.005$\tau$ and 0.01$\tau$ was used during crystallization and deformation process, respectively. An external pressure P=8 was applied, corresponding to the value of an atmospheric pressure.

**Deformation**

Uniaxial deformation was performed along the x direction of the simulation box under constant strain rate. This means the box dimension along the x direction changes linearly with time. The polymer chain was assumed to perform an affine deformation in accordance to the deformation of the simulation box. Firstly, the box size and shape is changed every time step, the coordinates of the particles are then updated. So the x positions of the particles are relocated proportional to the simulation box while the velocity is kept constant. The updated velocity and position of beads are then calculated accordingly. To investigate the structure/property relationship of semi-crystalline polymer under different conditions during deformation, three temperatures 0.2, 0.35, 0.7 and four strain rates $5\times10^{-6}$, $1\times10^{-5}$, $5\times10^{-5}$, $1\times10^{-4}$ were applied, respectively. The huge difference of structure transition and mechanical properties were found and will be discussed in the next section.

## Results and Discussion

### Crystallization

A molecular dynamics simulation of semicrystalline PE was performed in a static rectangular box with a side length ration of x:y:z=2:1:1. The ensemble is consisting of 200 coarse-grained chains with 500 repeating beads each chain, which is initially generated via a self avoiding random walk algorithm. A periodic boundary condition is applied in three dimensions to eliminate the boundary effect. During the simulation process, the temperature is controlled by a Nose-Hoover thermostat and a Berendsen barostat is applied to control the pressure. The initial conformation of the system is a non-equilibrium thermodynamic state and the conjugate gradient (CG) algorithm was performed to minimize the energy. The optimized conformation was then relaxed in an npt ensemble with the temperature of 0.1 and pressure of 16. Afterward the temperature of the system was increased to 1.0 followed by a relaxation process in a npt ensemble at atmospheric pressure. After sufficient relaxation, a well-equilibrated melting state with a disordered distribution of the molecular chains was generated. Subsequently, the isothermal crystallization process was performed after a sudden drop of the temperature from 1.0 to 0.7. In Fig 2, the snapshots of the morphology in the melting state and after crystallization are exhibited. In the melting state the polymer chains are randomly coiled, no ordered structure is observed. While after isothermal crystallization the system is composed of amorphous phases and crystalline phases, and the crystalline blocks are distributed randomly.

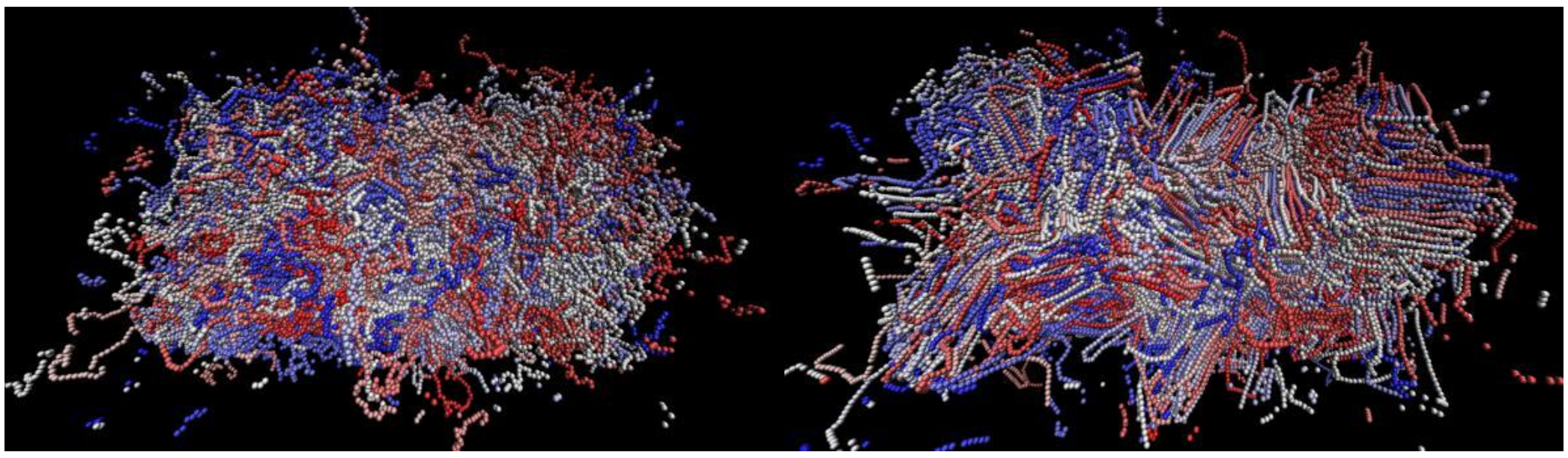

Fig 2. Snapshots of the melting state and the state after isothermal crystallization.

To assess the structure transformation during crystallization, the order parameters, S(t), and entanglement parameters are applied here. The order parameter is the most intuitive way to characterize the degree of order of the system. It is indispensable to characterize the emergence of crystal nucleus and the transition of the conformations from the coiled state to extension state. The total order parameter of the system is calculated as follows.[24],[25]

$$S(t) = \frac{3\langle \cos^2 \theta(t) \rangle - 1}{2} \tag{4}$$

Where *θ(t)* is the angle between the cord vector $r_{i+1} - r_i$ and $r_{i-1} - r_i$, where *i* is the center particles. The average is taken over all the angles in the simulation system. The evolution of order parameter during crystallization is represented in Fig 3. It is clearly that the order parameter exhibits a dramatic increase before the step reach to $5 \times 10^6$, and after that the increasing rate becomes slow down. The entanglement parameter is calculated following the atom steric methodology proposed by Yashiro et al.[26] The entanglement status of each atom is evaluated, by measuring the relative positions of the *k*th adjacent atoms to both directions along the polymer chain. If the angle between these two vectors is smaller than 90°, then the referred atom in the center is hereby designated as an entangled position. The entanglement density during crystallization process is shown in Fig 4. The entanglement density firstly increased with the steps which may due to the emergence of crystal nucleus. Subsequently, the entanglement parameter is decreased due to the rearrangement of polymer chains, which lead to transformation of the chains from the coiled conformation to the extend conformation. Latter, the entanglement parameter keeps constant due to the completion of crystallization. The angle distribution is showed in Fig 5 to clarify the evaluation of the conformation. It is obviously that three peaks are emerged in the figure which corresponding to three conformations in the bending potential. However, the position of the peak is not located rightly to the angle of 180 °, which correspond to trans-trans conformation in the bending potential. That is because the molecular chains in the system bear not only the bending potential, but also other force fields in the simulation system, for example, pairwise potential, bond stretching potential. It is the combined effects of all these force fields that determined the position of the peak. There is an obvious decrease in gauche-gauche and trans-gauche conformation while an increase in trans-trans conformation, which indicate crystallization from the melting state. To explore this process in depth, the trans conformation was defined as the angle larger than the threshold of 170 °. The evolution of trans-trans conformation during crystallization is represented in Fig 6. From the figure a similar tendency with the order parameter is observed, which indirectly support the rationality of these algorithms each other.

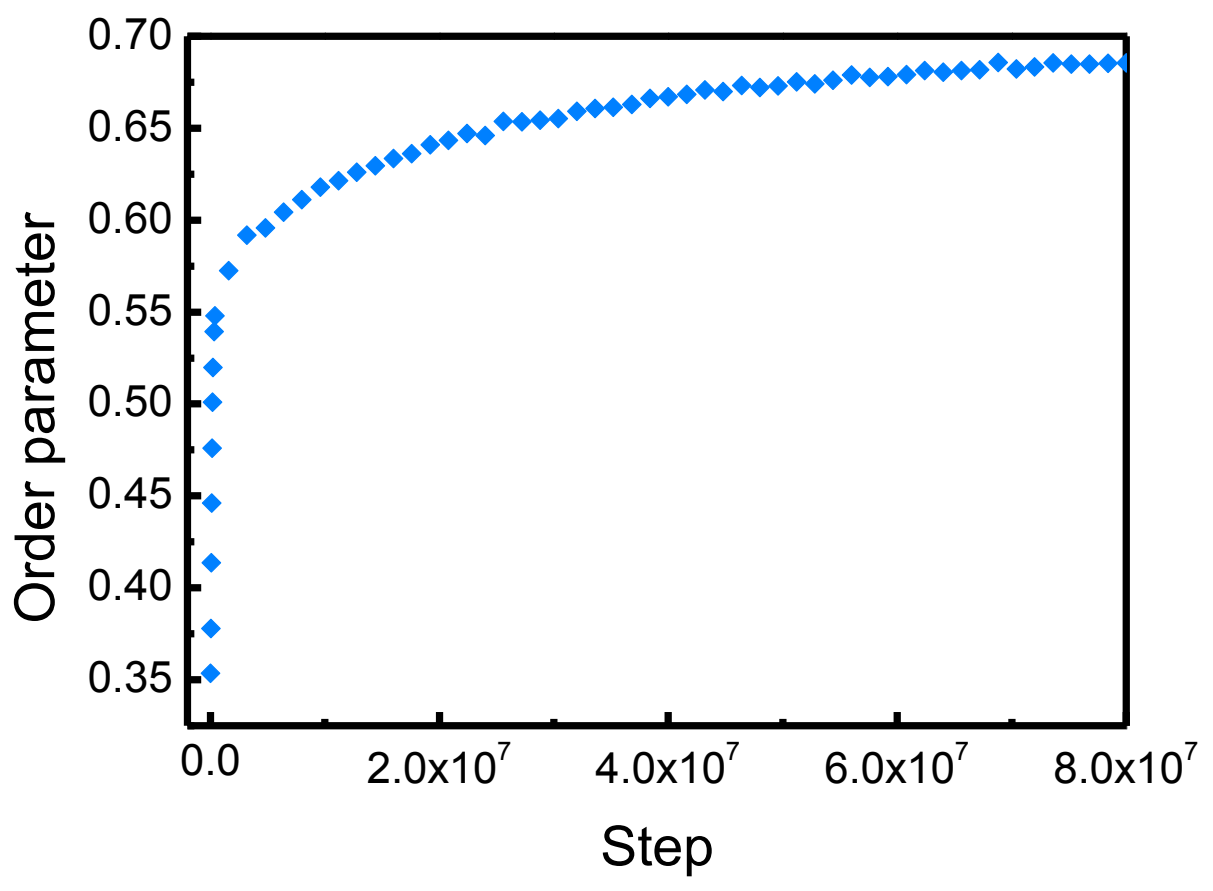

Fig 3. The change of order parameter during crystallization process

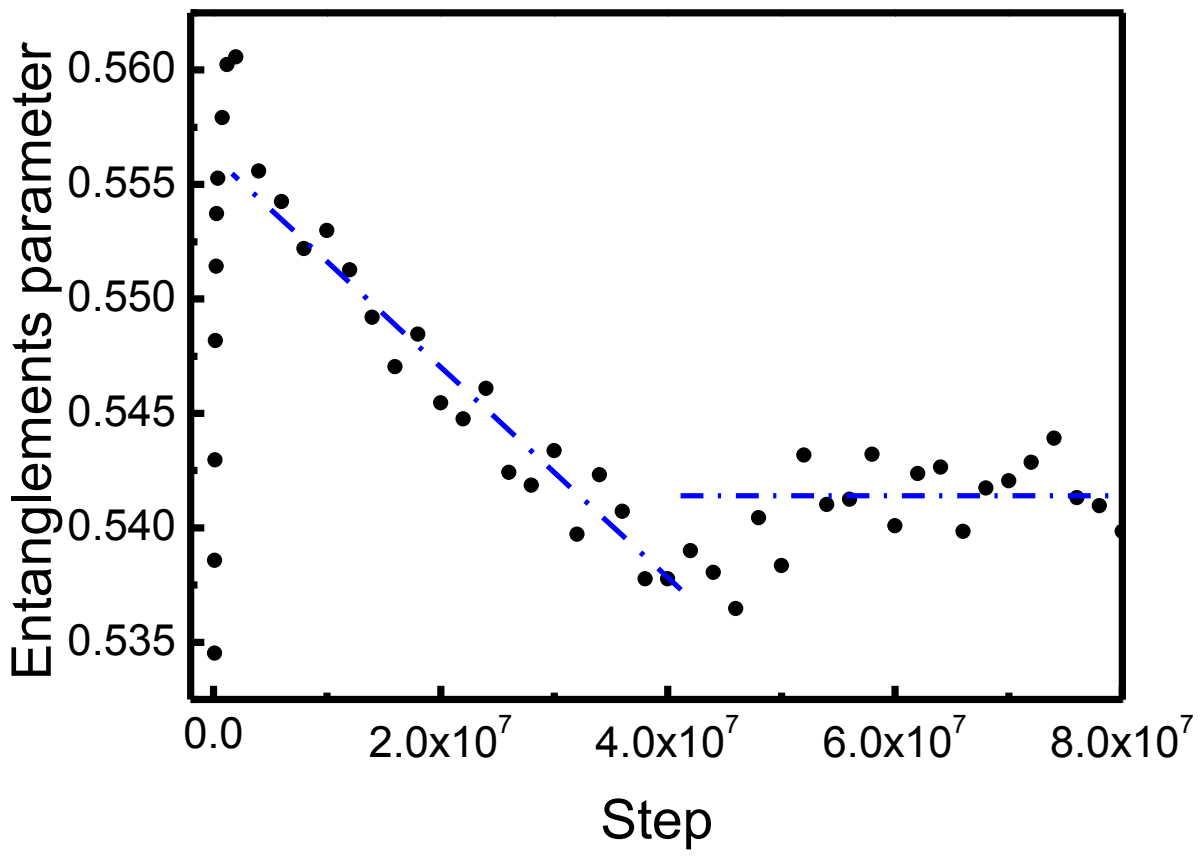

Fig 4. The change of entanglement parameter during crystallization process

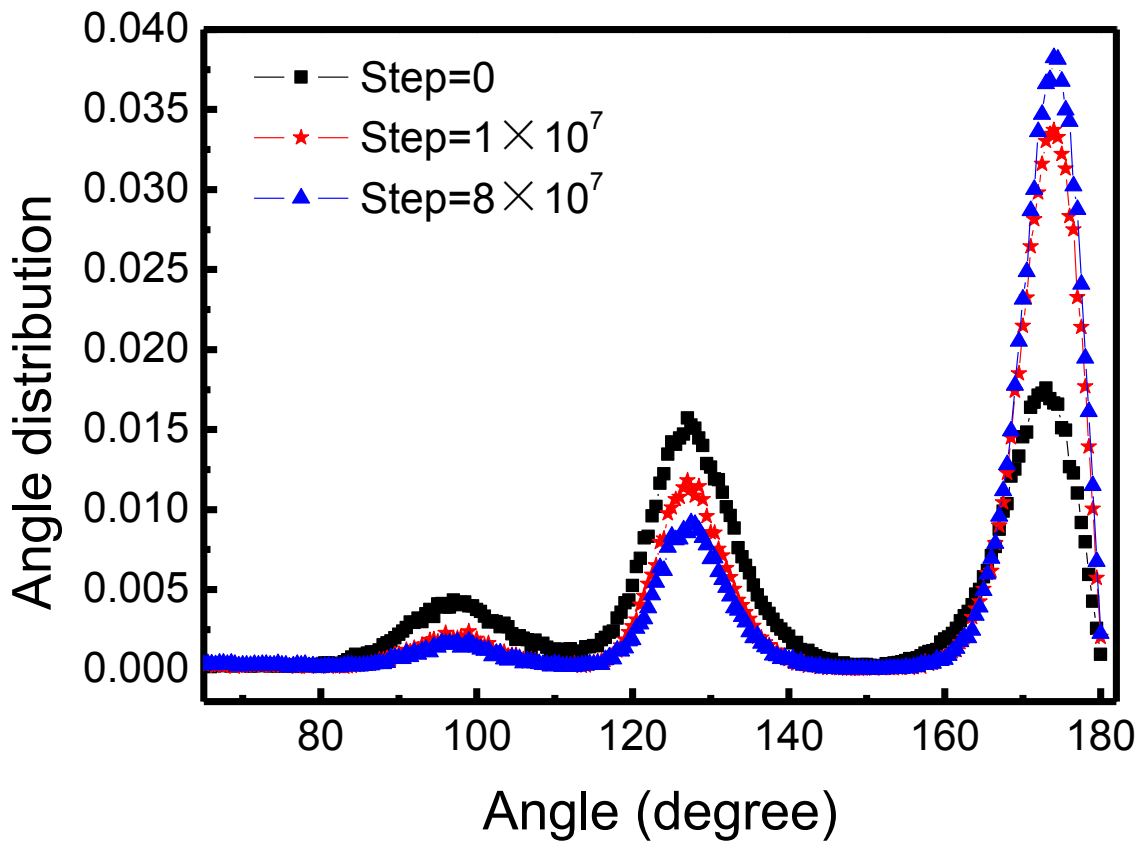

Fig 5. The angle distribution at different time steps during crystallization.

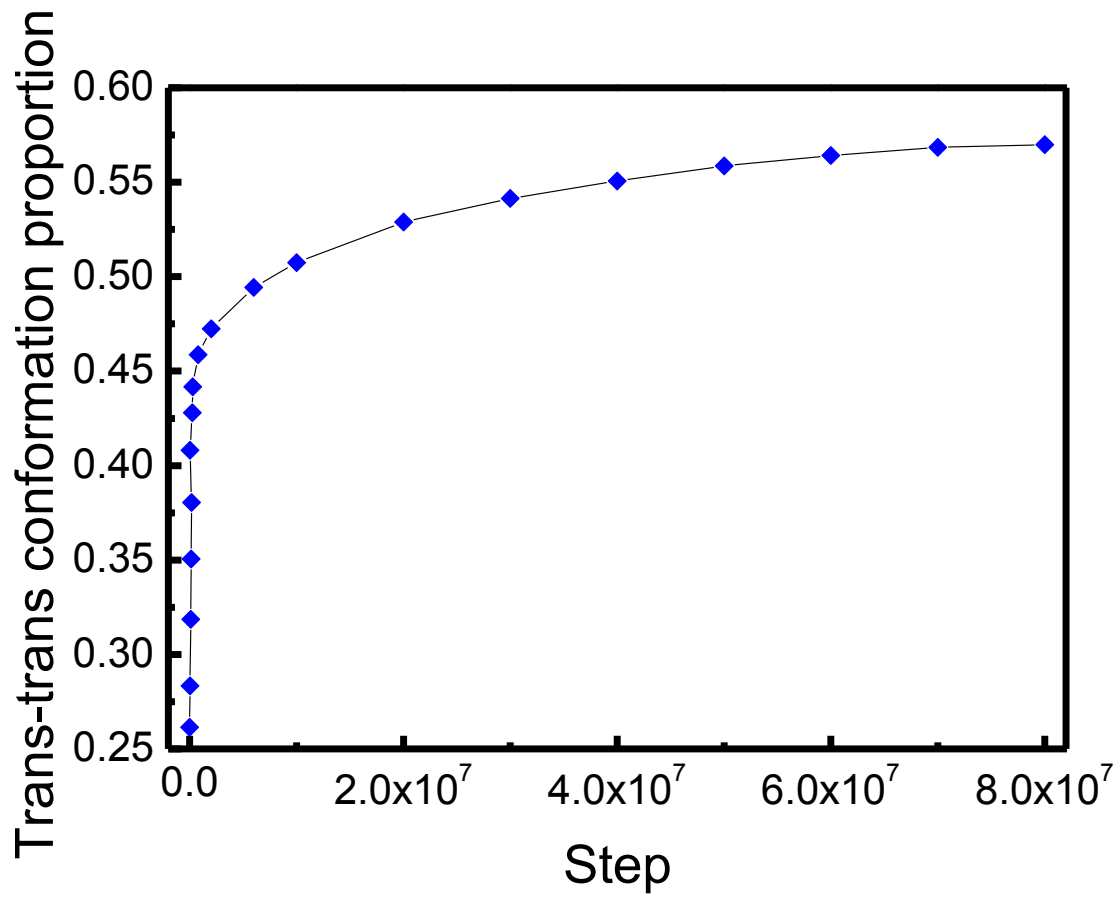

Fig 6. Trans-trans conformation proportion at different time steps during crystallization

## Equilibrium Crystallization Temperature, $T_{c0}$

Equilibrium crystallization temperature is an important parameter in studying the crystallization behavior. In this simulation, the crystallization temperatures are obtained by crystallizing with various cooling rates, as represented in Fig 7. By extrapolating the plot, the equilibrium crystallization temperature can be determined on the plot when cooling rate is zero. It can be seen from the figure that the $T_{c0}$ is about 0.785.

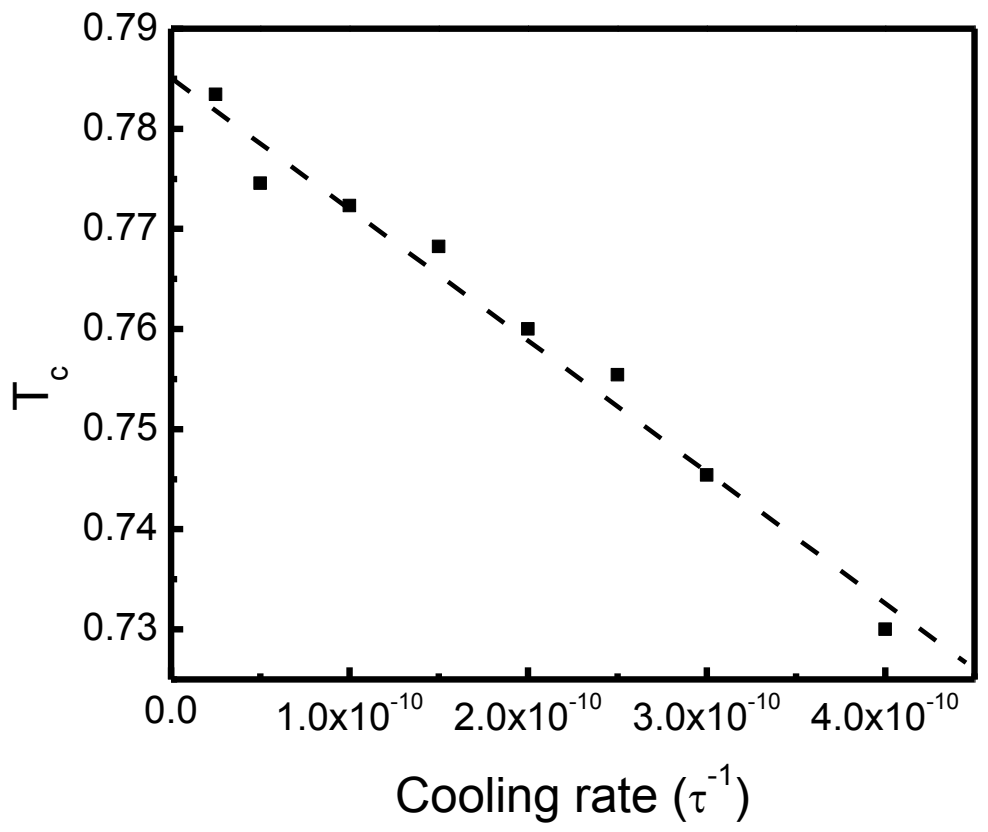

Fig 7. Crystallization temperatures at various cooling rates.

### Glass Transition Temperature, $T_g$

Glass transition temperature ($T_g$) is an important turning point in the chain segment's mobility. To obtain $T_g$, the melting state of the system was firstly quenched to a sufficient low temperature of 0.1, to make sure that no crystal lamellae was generated. Subsequently, a heating process was applied at a certain heating rate and the evolution of volume with temperature is represented in Fig 8. Two tangent lines were drawn from both ends and the intersect of the two lines was projected on X axis to estimate the glass transition temperature. The value of $T_g$ of the ensemble (200 chains with 500 DP) is measured as 0.349.

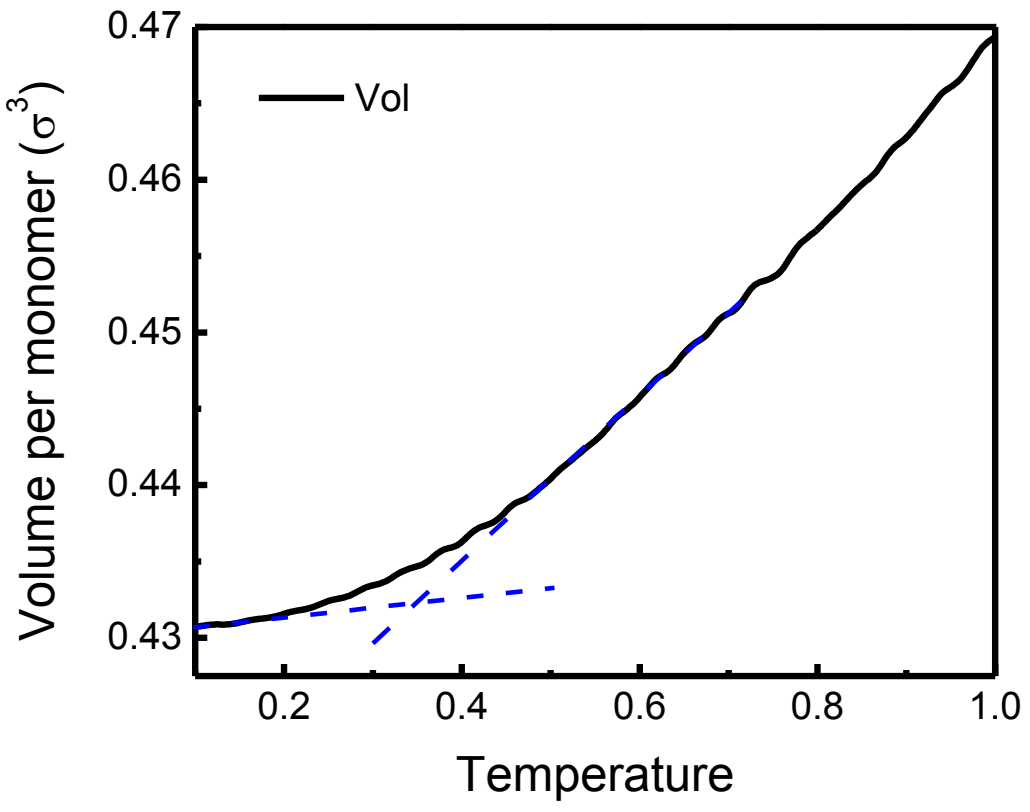

Fig 8. The evolution of volume with temperature

**Influence of chain length**

To investigate the influence of the molecular chain length to mechanical properties, two distinct ensembles were constructed, both have $10^5$ coarse-grained particles but with different chain length (200 chains with 500 DP and 100 chains with 1000 DP). Both ensembles have experienced the same isothermal crystallization process. The stress-strain curves and entanglement parameters at different temperatures are illustrated in Fig 9. Comparing the stress-strain curves, larger fluctuation was found in Fig 9(a). That was because a higher temperature T=0.7 was applied during deformation, which provided a high kinetic energy and a high mobility of the chain. At the high temperature of 0.7, the yield stress in the system with short chains was larger than the system with long chains. However, there emerges a conversion at the late stage of strain hardening regions. Strangely, at the low temperature of 0.2, the stress of the system with long chains is larger than the system with short chains in the whole deformation process. A key prerequisite should be understood firstly before clarify this phenomenon. Cause of the influence of chain length, a low crystallinity and a high entanglement density were processed in the system with long chains. At the high temperature of 0.7, both systems have enough kinetic energy and both chains have strong mobility regardless of chain length. In this condition, the amorphous phase is in a rubbery state. In the elastic region, the influence of the chain length and entanglement density is low, and crystallinity is the dominant factor. However, in strain hardening regime, more and more chain segments stretched along the deformation direction. The influence of the entanglements becomes more and more severely which resist the strain for further progress. This is because the more the entanglement points the more difficulty the mobility of the chain. When at the low temperature of 0.2, all the molecular chains have been frozen. The friction in the system with long chains is large, due to the interaction between long chains and high entanglement density. It is worth noting that the influence of crystallinity shouldn't be neglect. Comprehensively consideration, the chain length and entanglement density is the dominant factor in deformation mechanism at the low temperature. Summary from these two deformation mechanisms, it seems that there exists a critical temperature to distinct these mechanisms, according to the crystallinity and entanglement density which are determined by crystallization process and chain length.

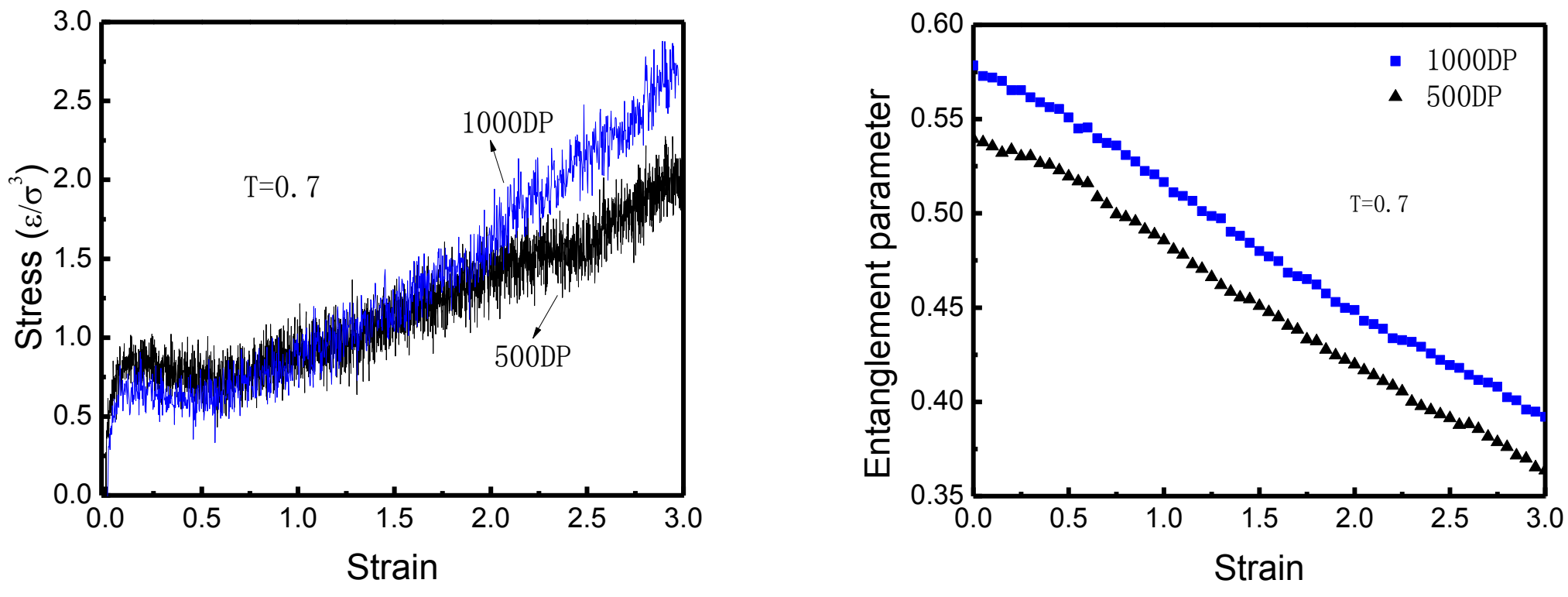

(a) (b)

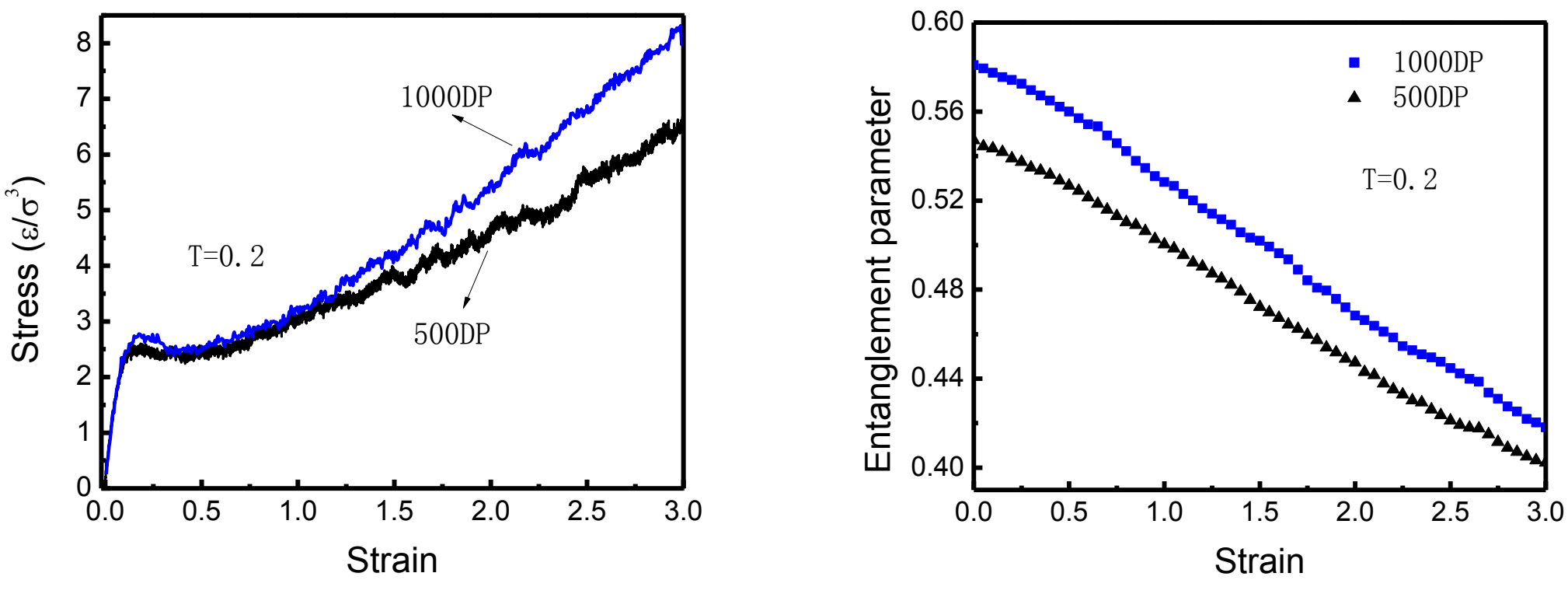

(c) (d)

Fig 9. Stress-strain curs of different chain lengths at temperatures T=0.7 (a) and T=0.2 (c).Entanglement parameters of different chain lengths at temperatures T=0.7 (b) and T=0.2 (d).

## Effects of strain rate and temperature to mechanical properties

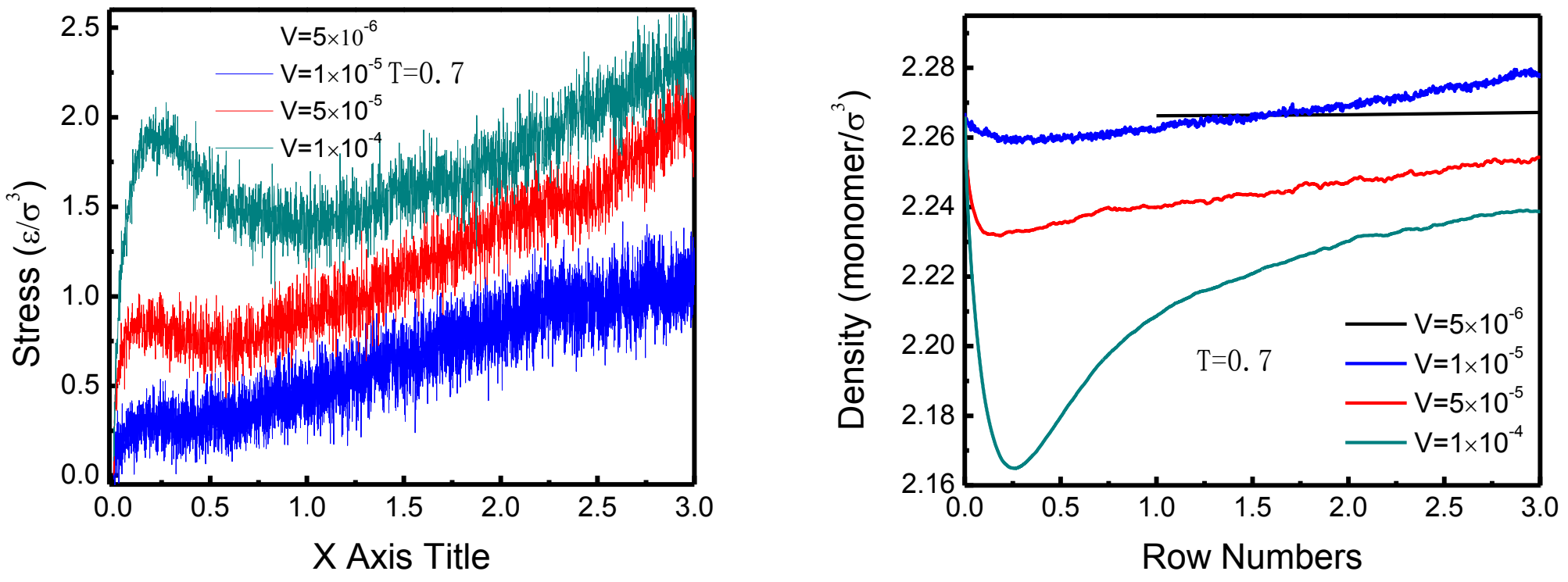

(a) (d)

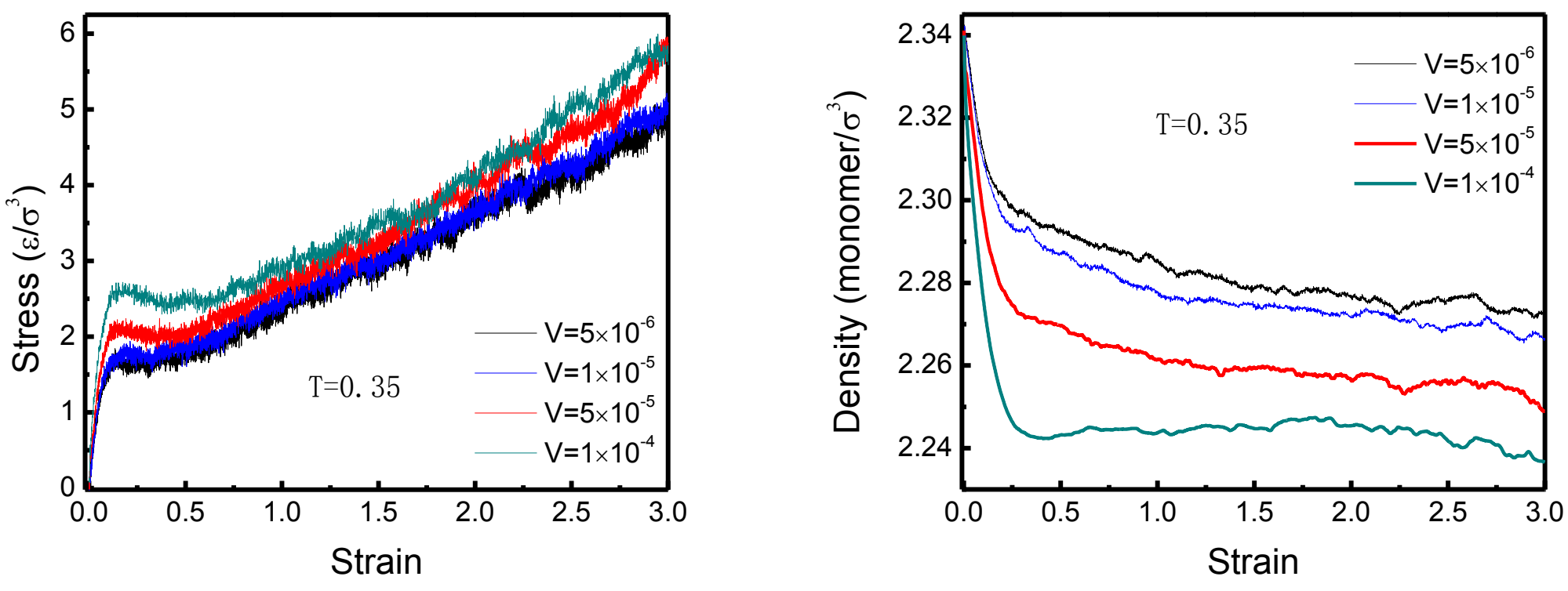

(b) (e)

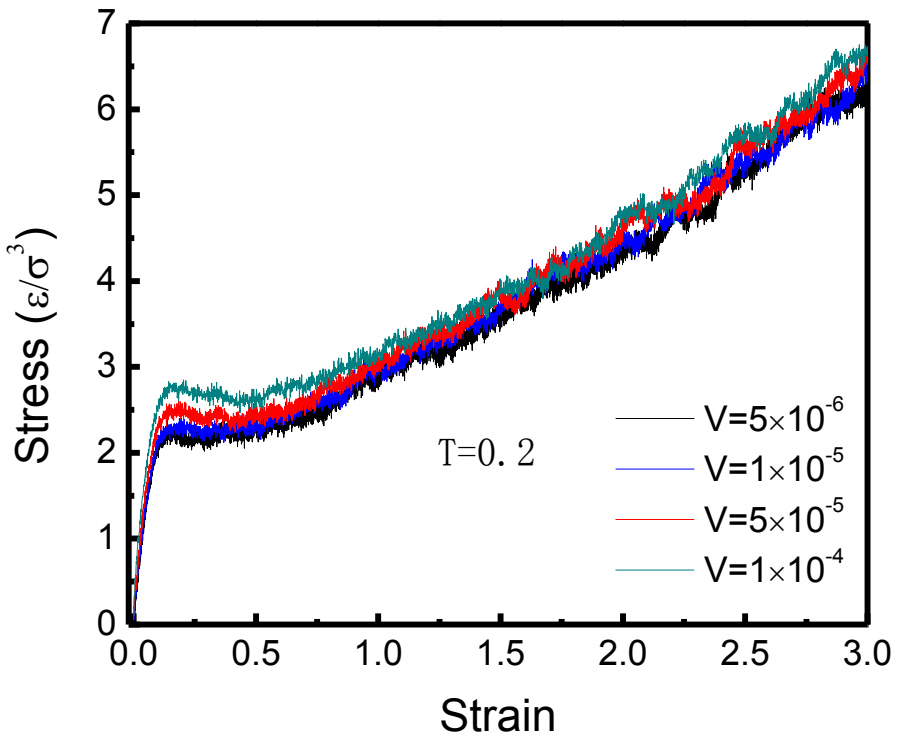

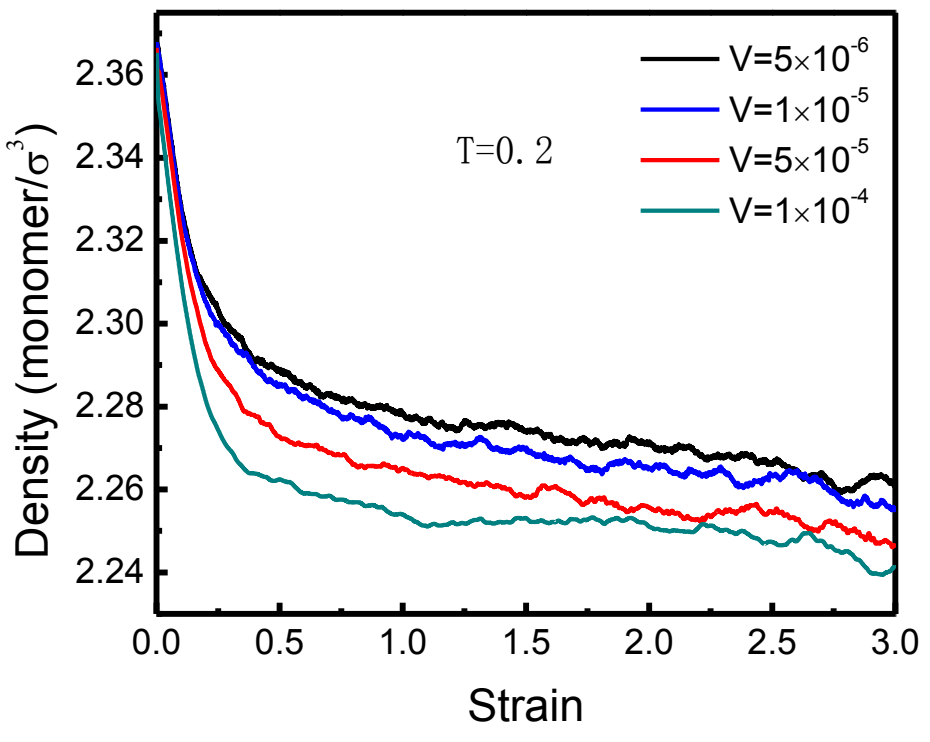

(c) (f)

Fig 10. Stress-strain curves and corresponding evolution of density at various strain rates at T=0.7 (a),T=0.35 (b). and T=0.2(c).The evolution of density at various strain rates at T=0.7 (d),T=0.35 (e). and T=0.2(f).

To investigate the mechanical responds of semi-crystalline polymers, a lot of deformation tests have been performed at various temperatures and strain rates. The stress-strain curves are represented in Fig 10. Both the cures exhibits elastic deformation, yielding and strain hardening behaviors. At relatively slow strain rates, $5\times10^{-6}$,$1\times10^{-5}$ in Fig 10(a), the stress-strain curves showed no strain softening regime after the yield point, following the strain hardening process directly. However, at the temperature of 0.2, the stress-strain cures at these two strain rates show a stress plateau after the yield point until the initiation of strain hardening.

At the high temperature of 0.7 and slow strain rates, the chain's mobility is strong, the amorphous phase is in a rubbery state. Due to the chain's strong mobility, the polymer chain can be easily orientated toward the deformation direction. After yield point, the crystal tilting and crystal lamellae slipping toward the stretching direction were happened, which lead the crystal stems orientated toward the deformation direction. Due to the high chain's mobility, the transition was progressed very quickly. In this stage the unfolding of the chain was not observed before the strain reach to 0.5. And the strain induced recrystallization toward the deformation direction was also observed latter at the interface between crystalline domain and amorphous domain. All in all it is the orientation of crystal stems that initiate the strain hardening behavior after yield point. At the low temperature of 0.2, the friction between the polymer chains is large. After yield point the crystal tilting and slipping toward the deformation direction was also happened but with a low transition rate. The crystal stems orientated toward the deformation direction may increase the stress. In the stress plateau stage, no crystal broken was observed. The unfolding progress was partially happened which may decrease the stress. Comprehensively consideration, the stress plateau may occur after the yield point at the slow strain rate.

The yield stress is increased with the increase of strain rate and temperature. This phenomenon can be clarified from the chain's mobility. When deformed at a high strain rate, the molecular chain's mobility couldn't come up with the change of strain, the friction between the molecular chains will be more larger. Thus, when reaching to the yield point the corresponding yield stress will be larger than at the small strain rate. The influence of temperature to the chain's mobility is similar to the strain rate. At low temperature, the intermolecular and intramolecular movement will be resisted, to make it yield, a more larger stress should be applied. From the figures we can conclude that the temperature and strain rate have played an important role in determine the yield stress.

From Fig 10(d), Fig 10(e) and Fig 10(f), a sudden drop process was happened in the density of the system before yielding in both three temperatures. It is obviously that the Poisson's ratio of the system is not 0.5. At the high temperature of 0.7, all the systems' density are then increased after the yield point. However, at the low temperature of 0.2 and 0.35, the density of the system declined all along the deformation process, but with a more slowly decline rate after the initial drop. These two different behaviors may due to the influence of temperature. In the elastic deformation, the crystal blocks remain intact, the deformation of the system was almost generated in the amorphous regions. Along the extension direction, the deformation was increased linearly with time, but the lateral contraction was too small at the elastic regime, which lead to the increasement of system volume. Based on this, the density was decreased with strain. After the yield point, the strain induced crystal tilting and slipping toward the stretching direction were happened. At the high temperature of 0.7, the mobility of the chain is strong, so a fast deformation of the crystal blocks toward the extension direction, which lead for further lateral contraction. And at the latter of strain hardening region, the strain induced crystallization along the extension direction may happen, which also lead to an increase of density. Subsequently, the density is increased after the yield point. However, at the temperature of 0.2 and 0.35, the chain's mobility is confined, which lead to a slow decline rate of the crystal blocks. In this case the crystal broken and unfolding of the chain is the main structure transition mechanism. It is hard for the recrystallization behavior from the amorphous regions to happen by the thermal motion of the molecular chains. Comprehensively consideration, the density is still decline but with a more slowly decline rate. From these two phenomena, it seems that there must be a critical temperature to make the density keep constant after the yield point. Another phenomenon was found that the density of the system was decreased with the increase of strain rate. This is because the slower the strain rate the smaller the resistance of the interaction between the chains. Thus, at the slow strain rate, the molecular chains have a strong mobility, which lead to the lateral concentration more easily. Subsequently, the system's volume will be smaller than the volume at the higher strain rate. So the density is larger at the relatively slow strain rate.

**Effects of strain rate and temperature to structure parameter**

*Bond length distribution*

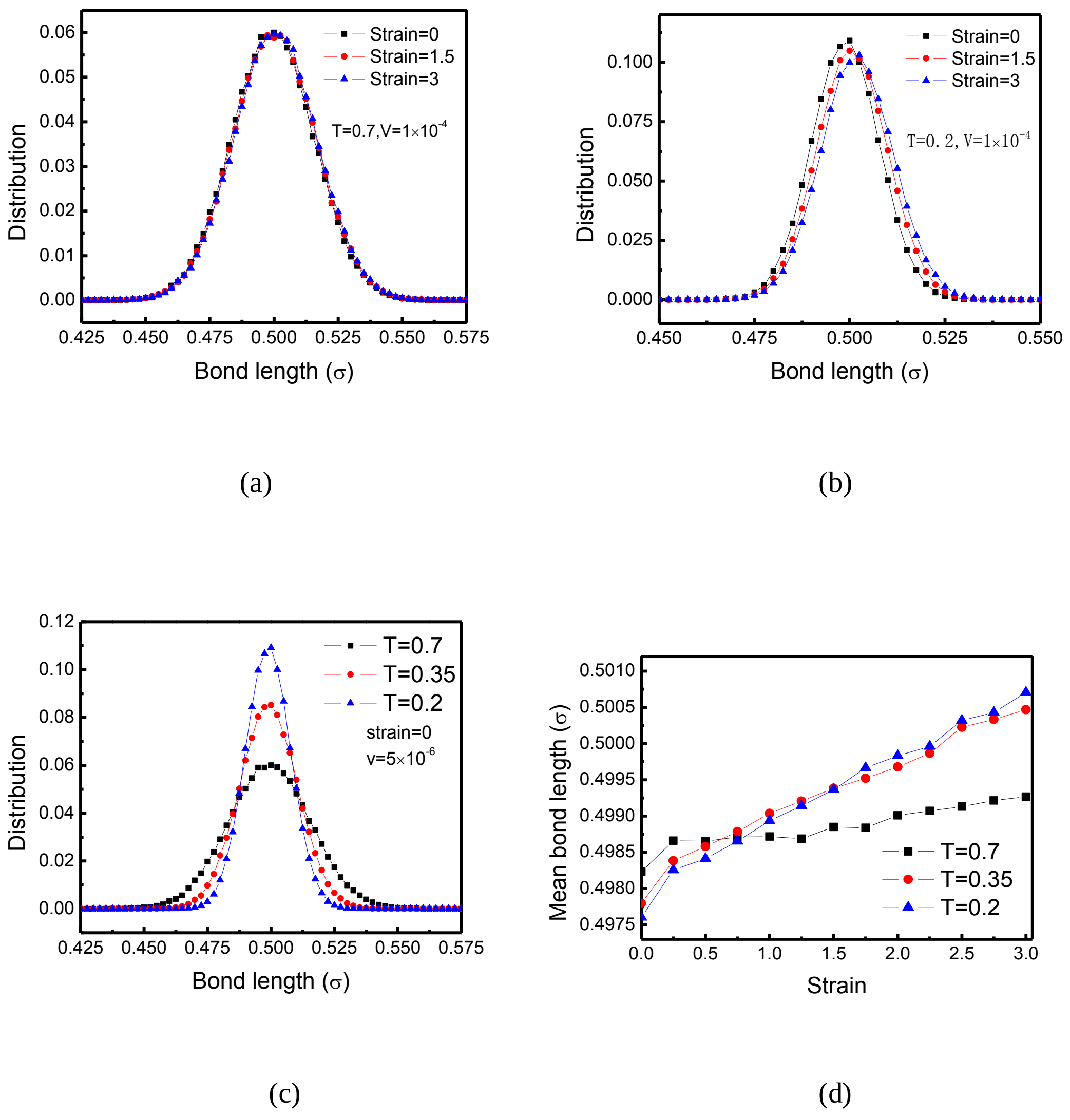

Fig 11. The distribution of bond length with strain at temperature T=0.7, (a); at temperature T=0.2, (b). The distribution of bond length with different temperatures at the unstretched state, (c). the evolution of mean bond length with stain at different temperature, (d).

The parameter of bond length distribution is an important method in characterize the structure transition during deformation, and the distribution of bond length with strain is represented in Fig 11. From the figures, at the high temperature of 0.7, the bond length distribution hardly changed with strain, while at the low temperature of 0.2, the bond length distribution represent a shift to the large values. This is because at the high temperature the molecular chain's mobility is also high, which lead to a short relaxation time of bond. While at the low temperature, the friction between the molecular chains is high, which lead to a long relaxation time.

Another manifestation is that the wider bond length distribution with the increase of temperature, and is represented in Fig 11(c). This is consistent with the increase of chain's mobility with temperature.

*Orientation parameter and entanglement parameter*

In order to characterize the extent of the chain stretching along the deformation direction, the orientation parameter was used. It was calculated using the Hermains' orientation function:

$$P_{orientation} = \frac{3}{2}\left\langle (\mathbf{e}_i \cdot \mathbf{e}_x)^2 \right\rangle - \frac{1}{2} \quad (5)$$

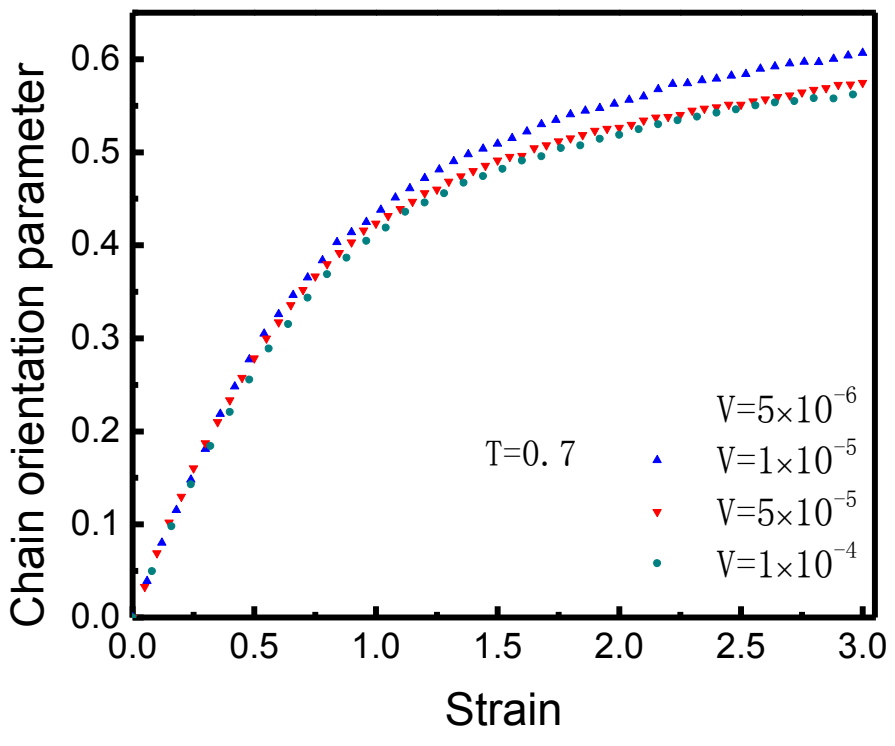

(a)

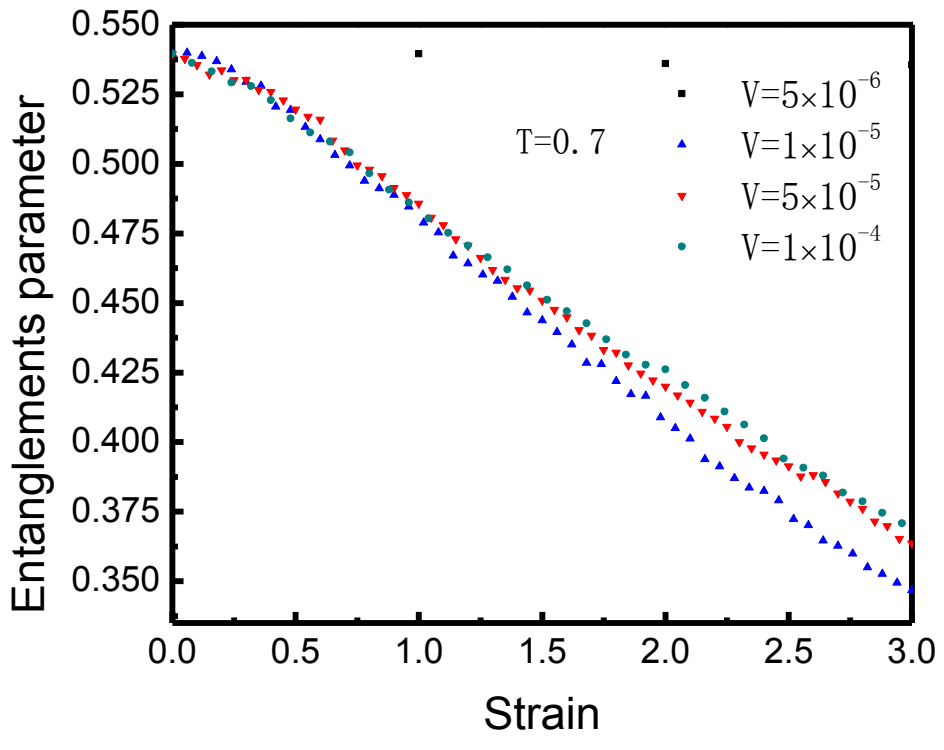

(b)

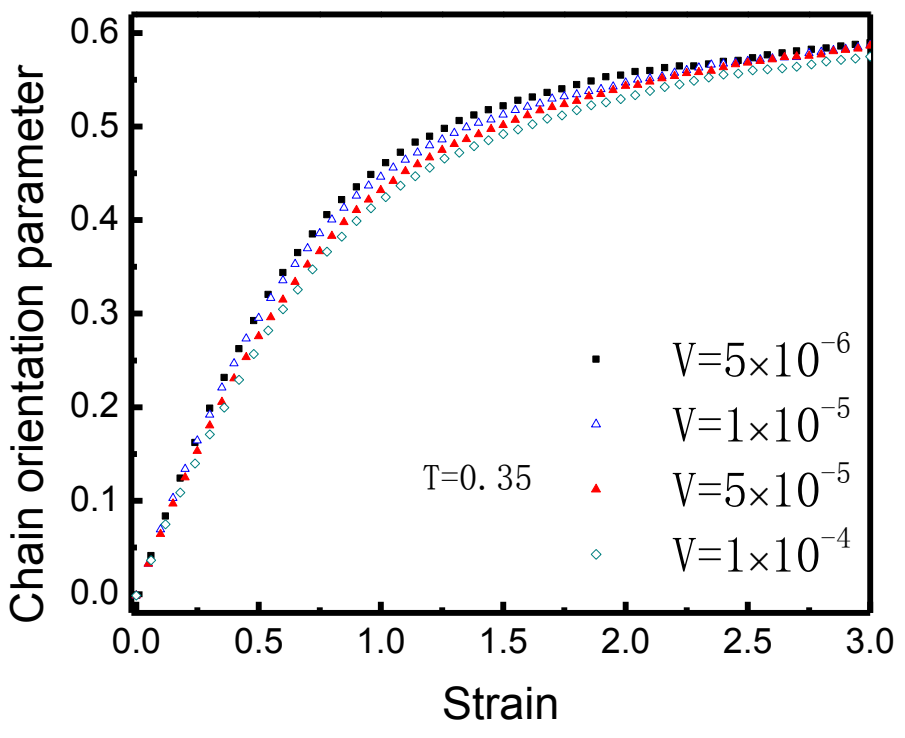

(c)

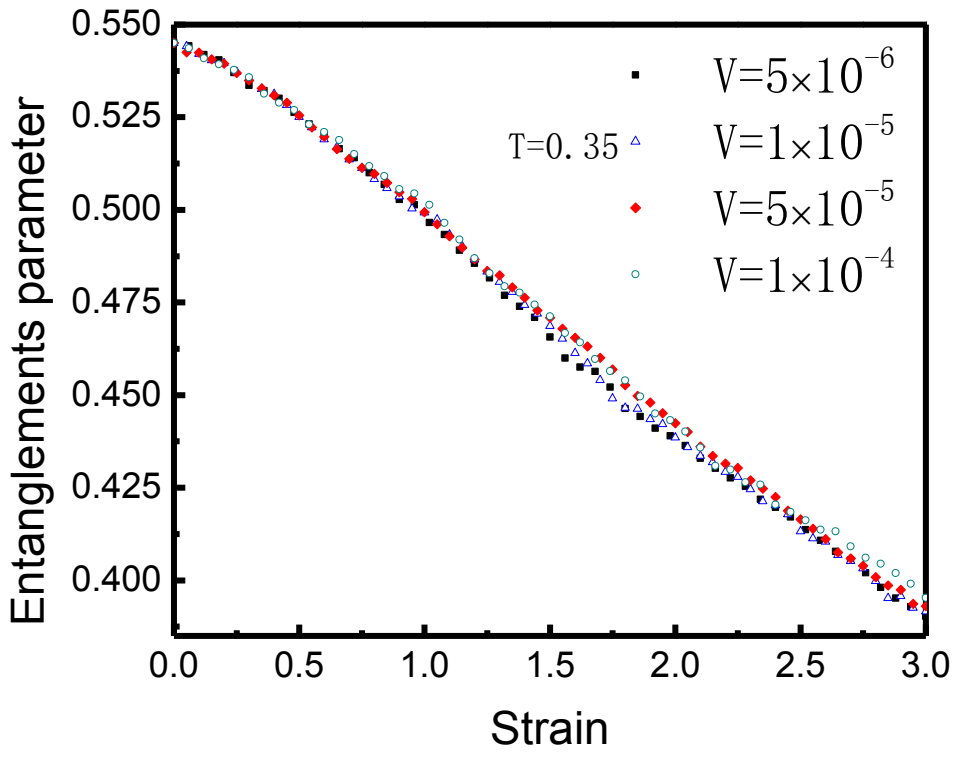

(d)

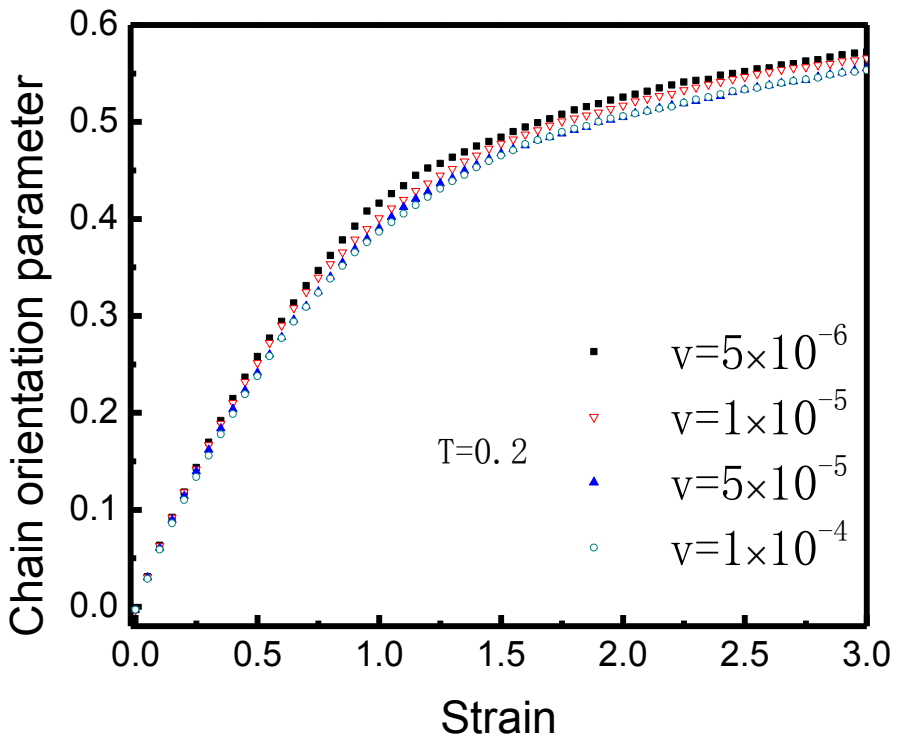

(e)

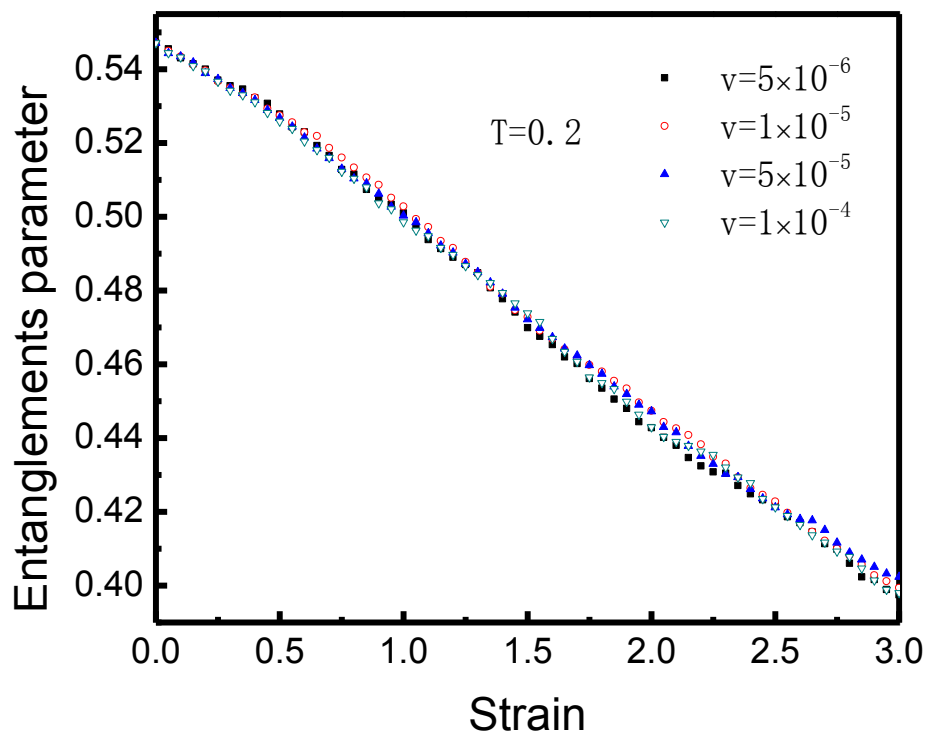

(f)

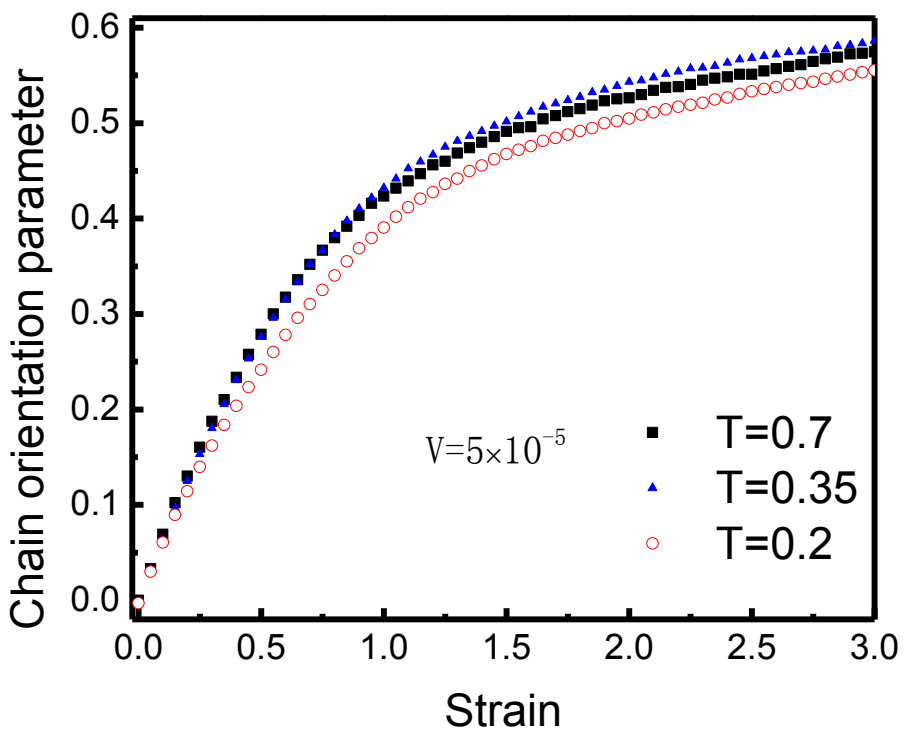

(g)

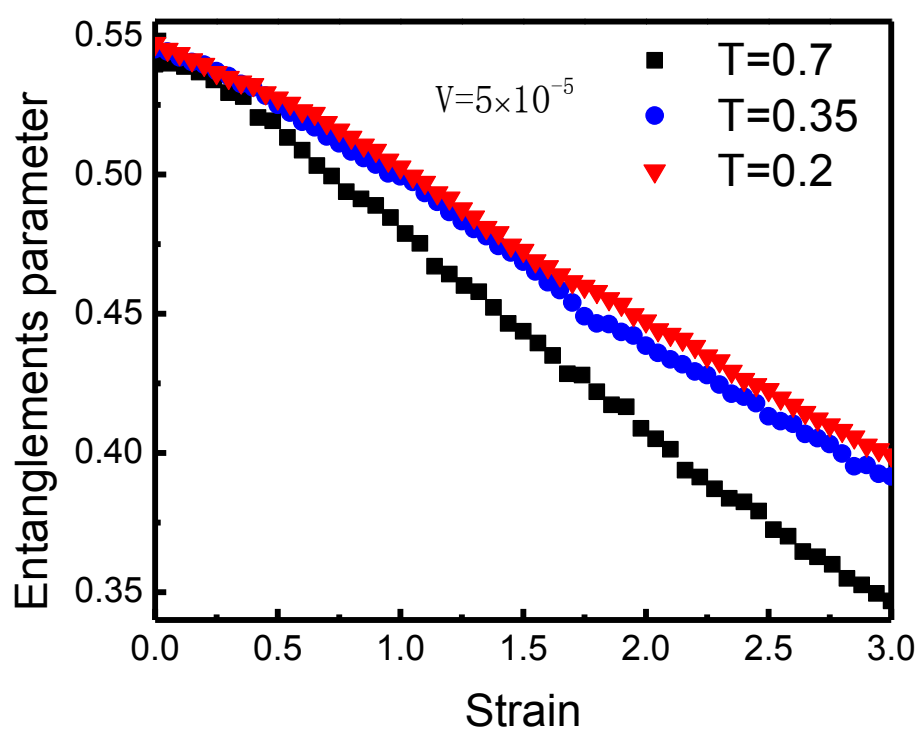

(h)

Fig 12. Evolution of orientation parameter and entanglement parameter during deformation (a), (b), (c), (d), (e), (f) at different strain rates and (g), (h) at different temperatures.

where $\mathbf{e}_i$ and $\mathbf{e}_x$ are the unit vectors characterizing the local chain direction and the extension direction, respectively. The angle bracket denote the average value over the whole simulation system. In this simulation a coarse-grained model with one bead represent a monomeric unit was employed. So the unit vector $\mathbf{e}_i$ was computed in the following method in this simulation.

$$\mathbf{e}_i = (\mathbf{r}_{i+1} - \mathbf{r}_i) / |\mathbf{r}_{i+1} - \mathbf{r}_i| \tag{6}$$

Since the uniaxial deformation was performed along the x-axis direction, therefore, the unit vector $\mathbf{e}_x = (1,0,0)$. The values of $P_{orientation}$ is varied between -0.5 to 1.0, which denote that the molecular chain is perpendicular to the stretching direction or parallel to the stretching direction, respectively.

The orientation parameters and entanglement parameters during uniaxial deformation at different temperatures and strain rates are represented in Fig 12. From these figures, the chain orientation parameters are all increased with the increase of the strain, while, the entanglement parameters have displayed an opposite trend. During uniaxial deformation, more and more chains will be aligned toward the extension direction which leads to the increase of orientation parameter. The disentanglement process was also occurred simultaneously due to the extension of polymer chains, which lead to the decreasing of entanglement parameter. From the figures the orientation parameters are increased with the decreasing of strain rate, while the entanglement parameters represent a decline tendency. This is because the chain will behaviors more flexible at the slow strain rate, and the molecular chain can be easily aligned toward the stretching direction, showing an increase in orientation parameter. This is also conducive to the progress of disentanglement behavior. Therefore, the entanglement parameter represents a decline tendency with the decreasing of strain rate. However, at the temperature of 0.2 and 0.35, from Fig 12(d) and (f), the entanglement parameter does not show a decreasement with the increase of strain rate. The cures are all winded together. It is believed that the temperature are too low, the speed of disentanglement at different strain rates becomes hardly to distinguish between each other.

During this simulation the influence of temperature is also considered. In order to investigate the influence of temperature to the evolution of microstructure during tensile test, three different temperatures were applied here. In Fig 12(h) the entanglement parameter is decreased monotonically with the increase of temperature. This is consistent with the phenomenon that the motility of molecular chain is increased with the increase of temperature. At the high temperature, the relaxation process can be easily performed, which for further promote the disentanglement progress. The influence of temperature to the orientation parameter represents some complicated relations. The orientation parameter is not increased monotonically with the increase of temperature and is represented in Fig 12(g). The orientation parameter at T=0.35 is larger than the value at T= 0.7. This anomaly phenomenon is also not the first time to be found. In fact, in our previous paper in investigating the deformation mechanisms of amorphous polymers[27] this phenomenon have been occurred. As is well known that the higher the temperature the more flexible the chain will be, that is to say the chain will be more easily aligned toward the stretching direction at high temperature during deformation. However, another influential factor couldn't be neglect, at the high temperature the kinetic energy is also high, the relaxation time of the chains is small, which will lead to the increase of the trans-gauche and gauche-gauche conformations, especially when the potential barrier between the trans conformation and gauche conformation is not too high. The evolution of gauche-gauche conformation trans-gauche conformation and trans-trans conformation with temperature are showing in Fig 13(a). From this figure, the trans-trans conformation is decreased with the increase of temperature, while the gauche-gauche conformation and trans-gauche conformation represent a rising trend. The evolution of trans-trans conformation with stain at different temperatures are also represented in Fig 13(b). During deformation, the value of trans-trans conformation at temperature of 0.7 is smaller than the values at the temperature of 0.35 and 0.2. Conclude from the above analysis, raising temperature have double influence on polymer chains during deformation, one is make the chain more flexible and easily be stretched, the other is reducing the relaxation time. Actually, the orientation parameter is determined by the interaction of these two effects under the certain strain rate.

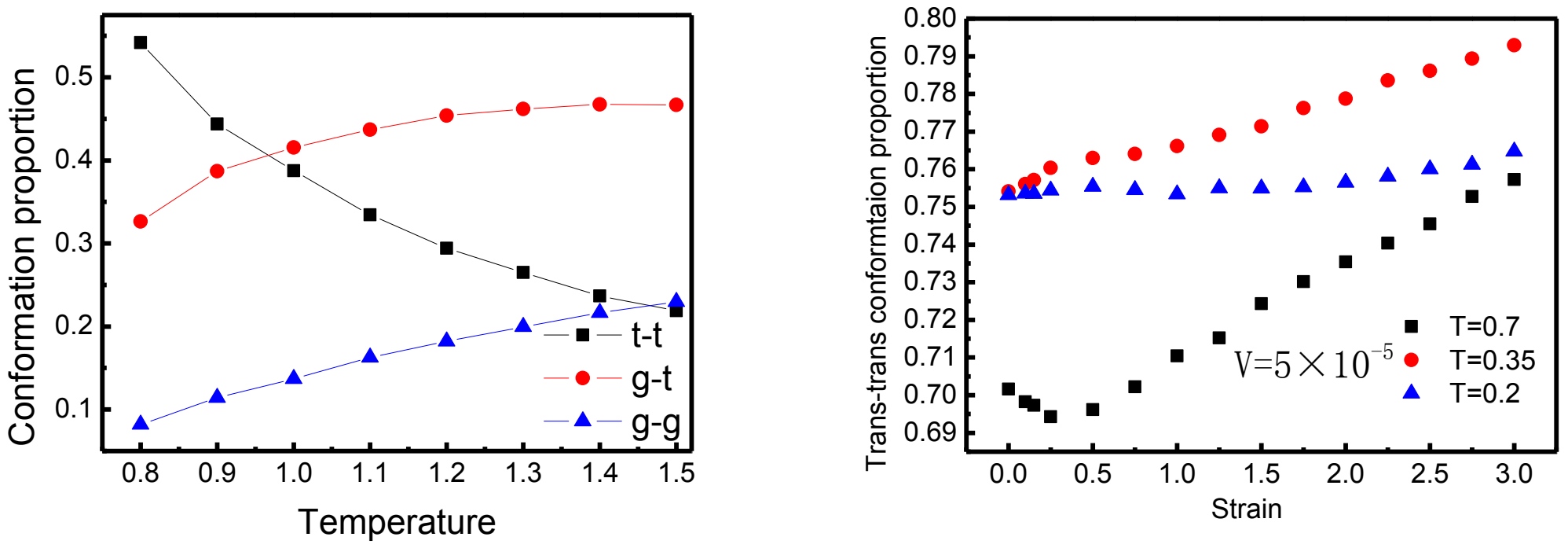

(a) (b)

Fig 13. Conformational changes with temperature (a), (b) the change of trans-trans conformations during deformation at different temperatures.

## Evolution of trans-trans conformation during deformation

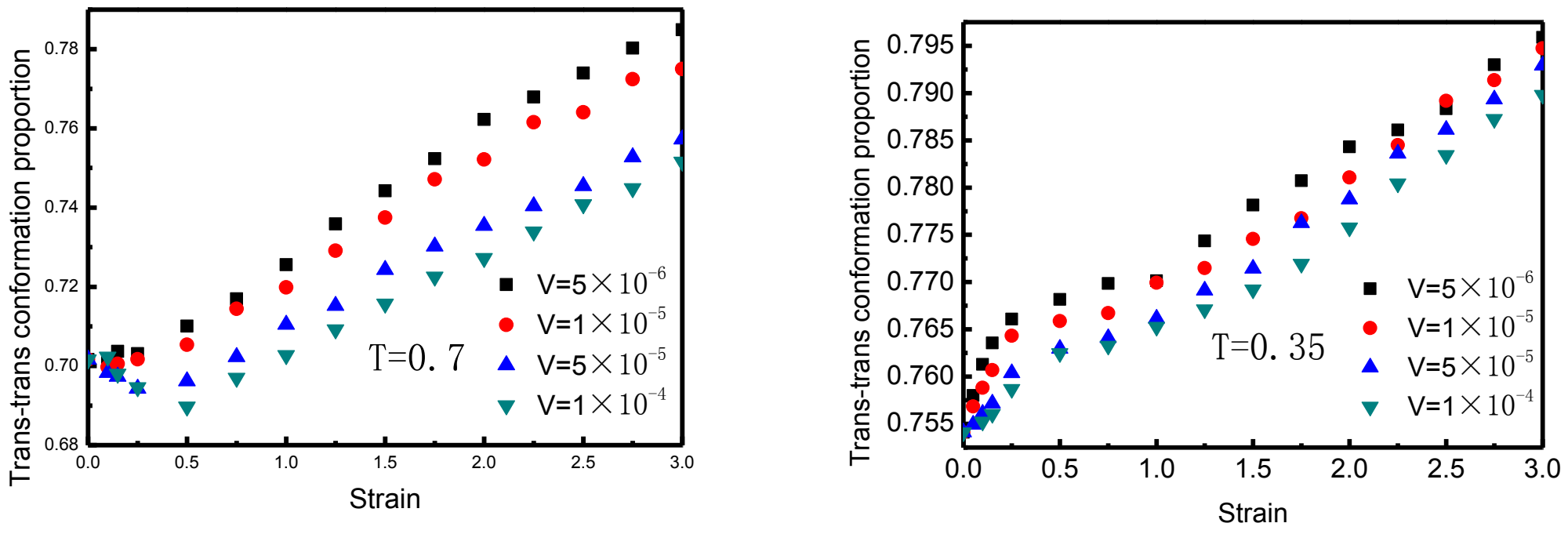

(a) (b)

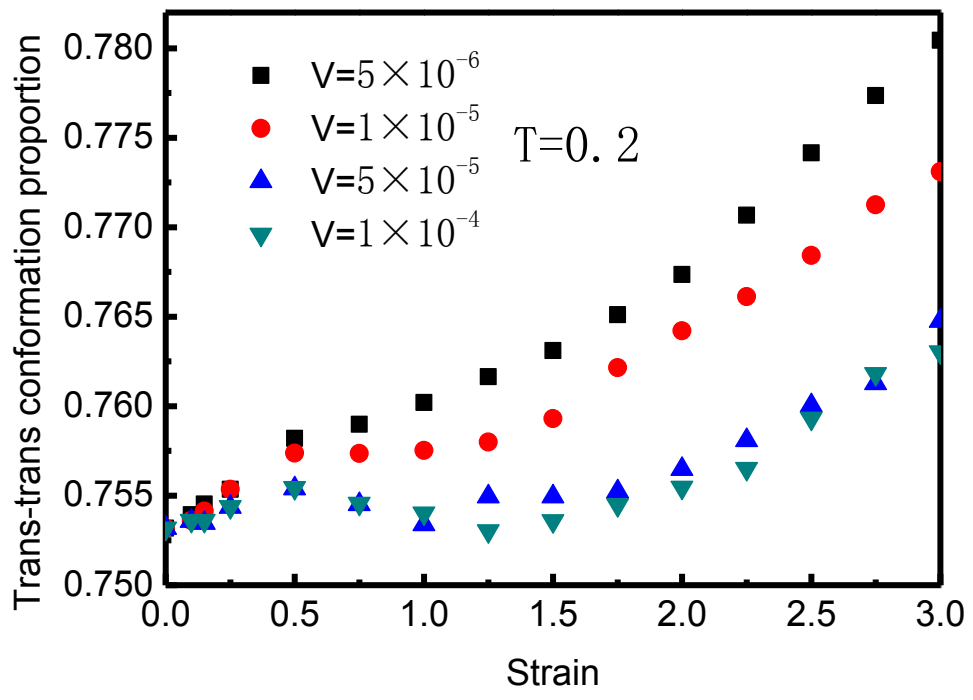

(c)

Fig 14. The evolution of trans-trans conformation with strain at different strain rates.

For subtle description in the structure evolution of semi-crystalline polymer in the process of deformation, the trans-trans conformation with the strain at various strain rates and temperatures are represented in Fig 14. From these figures, the proportion of trans-trans conformation in the system have experienced different process. At high temperature of 0.7, the trans-trans conformation proportion all experienced a decline at the initial deformation, and after the certain transition point the value of the proportion increased steadily with strain. However at the deformation of 0.35 and 0.2, the trans-trans conformation proportion all experienced an increase at the initial deformation.

At the high temperature of 0.7, the molecular chains have a strong mobility, and the stability of the crystalline is low, which may lead to partially unfolding process in crystal domain with the molecular chains go into the amorphous phase, thus the initial decrease of trans-trans conformation. At this temperature, the molecular chain can be easily aligned toward the stretching direction and the strain induce recrystallization process from the amorphous phase to form fibrillar structure can also be happened, and in turn, make the trans-trans conformation increase rapidly after the certain transition point. The snapshots of the system during deformation at different temperatures and strain rates are showed in Fig 15.

When at the low temperature of 0.35, the friction between molecular chain is very high, and the high stability of the crystalline. At the initial deformation the crystalline domain keeps intact, it is the orientation of the molecular chain in the amorphous regions that lead to the increase of trans-trans conformation. The crystal broken will be happened after the yield point, and let the molecular chain go into the amorphous regions, which will slow the increase rate of trans-trans conformation of the simulation system. At the temperature of 0.2, the mobility of the chain is more slower, thus, the increase rate of the trans-trans conformation at the initial is smaller than the value at the temperature of 0.35. At some rapid strain rates, $5\times10^{-5}$, $1\times10^{-4}$, the trans-trans conformation represent a slow decline after the initial increase. It is the influence of two aspects. For one aspect, the temperature is too low and the strain rates are high, due to the large friction between molecular chains, the trans-trans conformation transition is progressed very slow. For the other aspect, it is easily for the broken of the crystal domain at these conditions, and then reduce the trans-trans conformation. It is the interaction of these two aspects that makes the decline of the trans-trans conformation after the initial increase. Subsequently, the trans-trans conformation will be increased due to the further orientation of the chain toward the deformation direction. The other phenomenon is found in Fig 14 that the trans-trans conformation is larger at the slow strain rate than at the high strain rate along the deformation process. This is because the high the strain rate the slow the chain’s mobility and then the small proportion of trans-trans conformation during deformation.

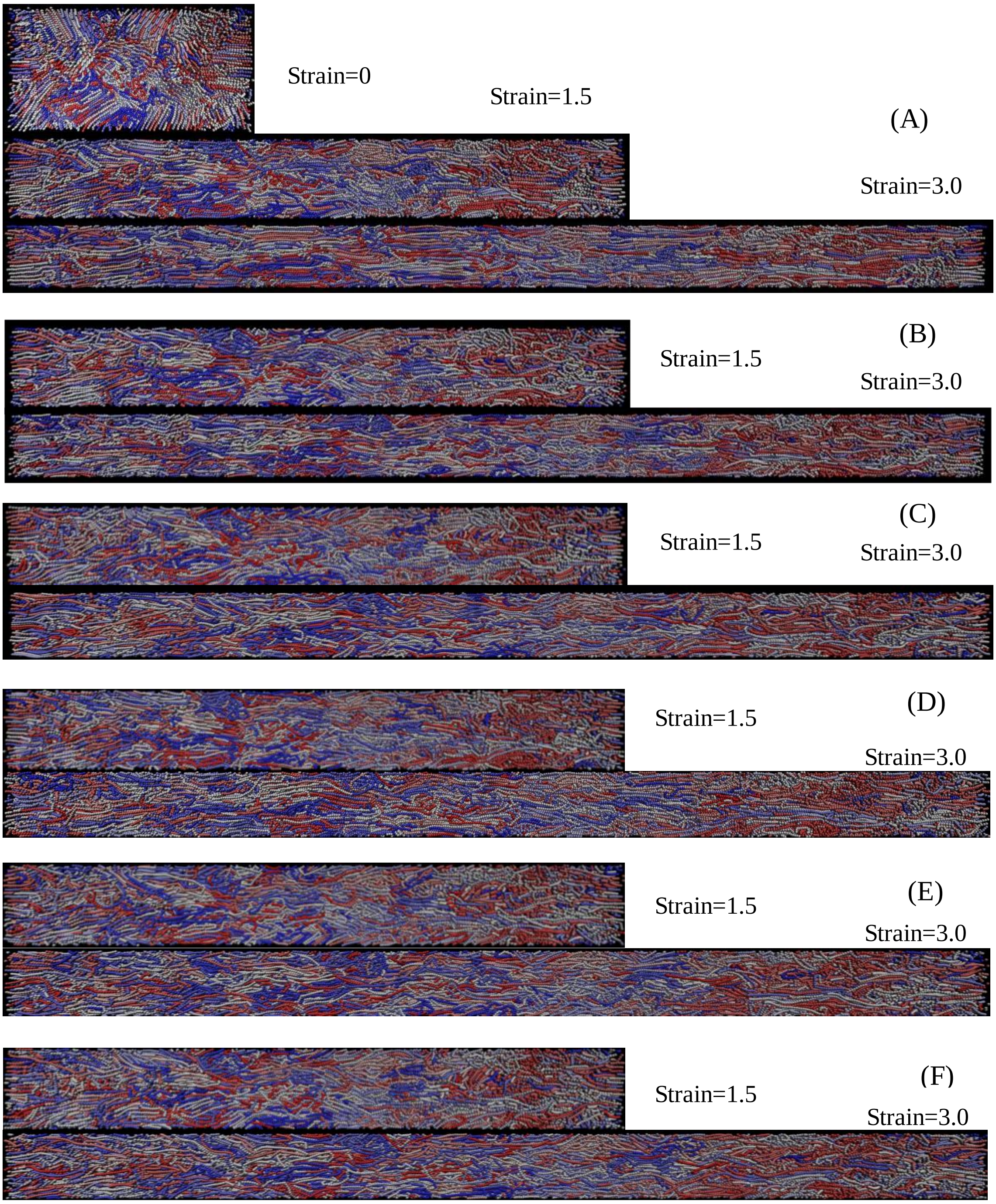

Fig 15. Snapshots of the simulation system during unaxial deformation at the strain rate of $5\times10^{-6}$ and temperature of T=0.7, (A); T=0.35, (B); T=0.2, (C). And strain rate of $1\times10^{-4}$ at the temperature of T=0.7, (D); T=0.35, (E); T=0.2, (F), respectively.

## Conclusion

A coarse-grained model of semicrystalline PE was built through isothermal crystallization by molecular dynamics simulation to investigate the correlation of mechanical properties and microstructures during uniaxial deformation. The influence of chain length to structure transition and mechanical properties is worked through two aspects. For one aspect, the longer the chain length the smaller the crystallinity of the ensemble. For the other aspect, the longer the chain length the larger the entanglement parameter and friction between the molecular chains. At the high temperature before yielding, the crystalinity have played an important role in the mechanical properties. At the low temperature, the friction between molecular chains and entanglement parameter have played an important role in mechanical properties in the whole deformation.

Different strain rates and temperatures have been applied to the simulation system to investigate the effects on the mechanical properties. The yield stress is increased with the increase of strain rates and falling of temperatures. During deformation all the systems' density represent a decreasement at the initial regardless of the temperatures and strain rates. However, at the high temperature, the density will then increase after the minimal point, while at the low temperature, the density was still decreasing but with a more slowly decline rate. The orientation parameter of the system is increased with the decreasing of strain rates, while the entanglement parameter represent an opposite trend. A strange phenomenon was found that the orientation parameter is not increased monotonically with the increase of temperature. The temperature have double effects on the polymer chain during deformation, one is make it more flexible and easily be stretched, the other is reducing the relaxation time. Orientation parameter is determined by the interaction of these two effects under the certain strain rate.

During deformation the trans-trans conformation shows different evolution process at different temperatures. It is the interaction of the chain's mobility and the stability of the crystal domain that determine the mechanism of the conformation evolution. The evolution of the trans-trans conformation have a lot to do with the evolution of crystal blocks and the formation of fibril structures. It can be concluded that temperature have a great effects on the behavior of the semicrystalline polymers during deformation. And increasing strain rate has some similarity effects on structure change with lowing temperature. Deeply understanding the effects of temperature and strain rate on the chain's mobility and stability of the crystalline region are an important bridge to reveal the structure/property relations of semi-crystalline polymer during tensile deformation.

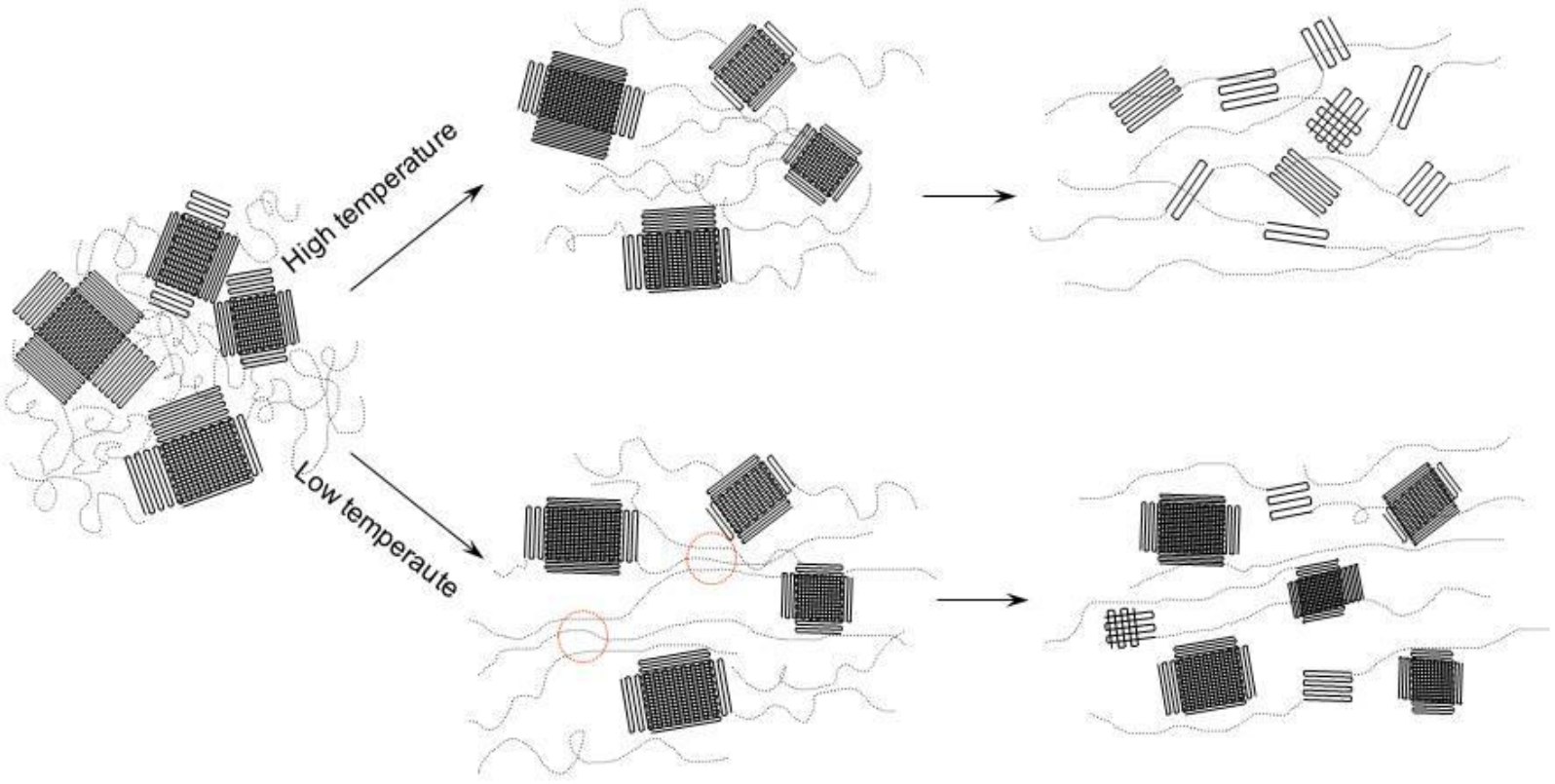

Fig 16. The mechanism of the structure changes in the process of deformation at different conditions.

## Acknowledgment

The authors appreciate the financial support from the National Science Foundation of China on this study.